\documentclass[11pt,letterpaper,journal,oneside,onecolumn]{IEEEtran}
\usepackage{cite}
\usepackage{graphicx}
\usepackage{epsfig}
\usepackage{balance}
\usepackage{setspace}

\begin{document}
\thispagestyle{empty}
\pagestyle{empty}

%\begin{comment}

%For double coloumn....
\title{A Radial-Dependent Dispersive Finite-Difference Time-Domain Method for the Evaluation of Electromagnetic Cloaks}

\def\affil #1 {\begin{itemize} \item[] #1 \end{itemize}}

\author{Christos Argyropoulos,
        Yan~Zhao,
       Yang~Hao,}
\maketitle

\affil{Queen Mary, University of London, Mile End Road, London, E1 4NS, United Kingdom\\
Fax: + 44-20-78827997; email: christos.a@elec.qmul.ac.uk}

%\end{comment}

%\title{A Radial-Dependent Dispersive Finite-Difference Time-Domain Method for the Evaluation of Electromagnetic Cloaks}
%\author[org1]{C. Argyropoulos*}
%\author[org1]{Y. Zhao}
%\author[org1]{Y. Hao}

%\address[org1]{Electronic Engineering, Queen Mary University of London\\Mile End Road, London E1 4NS, UK\\E-mail: christos.a@elec.qmul.ac.uk}

%\maketitleblock

\doublespace

\begin{abstract}
A radial-dependent dispersive finite-difference time-domain (FDTD) method is proposed to simulate electromagnetic cloaking devices. The Drude dispersion model is applied to model the electromagnetic characteristics of the cloaking medium. Both lossless and lossy cloaking materials are examined and their operating bandwidth is also investigated. It is demonstrated that the perfect ``invisibility" from electromagnetic cloaks is only available
for lossless metamaterials and within an extremely narrow frequency band.
\end{abstract}

\section{Introduction}
In the last two years, cloaking devices received unprecedented attention from the scientific community. Linear coordinate transformations have been applied in order to manipulate the electromagnetic characteristics of the propagation medium \cite{Pendry,Leonhardt,Leonhardtnotes}. These techniques originate in the theory of General Relativity and conformal mapping procedures. After the transformations, the medium which is produced, the cloaking shell, is able to guide the electromagnetic waves around an object without any disturbances and reflections. This is like the waves are propagating through the free space. Hence, the object is practically ``invisible'' to an exterior viewer. The permittivity and permeability of such a cloaking shell are anisotropic and dispersive as was first demonstrated by Pendry \textit{et al.} \cite{Pendry}.

The most appropriate materials for the production of the cloak's electromagnetic characteristics are the recently discovered metamaterials \cite{Veselago}. They are artificial materials with extraordinary electromagnetic properties which cannot be obtained in nature. The cloaking device is easier to be implemented in practice when not all the parameters are radial dependent. To solve this problem, reduced parameter sets were proposed in \cite{Cummer,Cai,Qiu} which operate at different polarizations. A simplified cloaking device was constructed and tested in microwave frequencies with promising results \cite{Schurig}. Currently, there are efforts to experimentally verify cloaking at optical frequencies \cite{Cai}. It is supposed to be constructed from silver nanowires with subwavelength dimensions embedded in a silica dielectric host. Another approach towards the implementation of an optical cloak is by using a concentric structure which is made of a layered gold-dielectric material \cite{Smolyaninov}. Recently, cloaks which are derived from a higher-order coordinate transformation \cite{Chettiar} were proposed for a future optical cloaking device \cite{M.Shalaev}. Indeed, it would seem that the scientific community is one step closer to the impossible achievement of ``invisibility'', which before could only have been part of a science fiction scenario.

However, metamaterials are dispersive which directly leads to limited frequency band capability. These limitations are thoroughly analyzed in \cite{Liang,Yao}. Another drawback of metamaterials is their lossy nature. Furthermore, the ideal cloaking realization is impossible in theory due to the wave nature of light \cite{Wolf}. To avoid using metamaterials, an alternative approach to construct the cloaking shell from layers of homogeneous isotropic materials with subwavelength dimensions was proposed in \cite{Huang}. However, it is difficult to realize the previously mentioned structure due to the alternating different values of permittivities which are required for the layers of the cloaking device. A different approach which applies sensors and active sources near the surface of the cloaked object, has been described in \cite{Miller} and it can be functional in a broader bandwidth. Finally, wider frequency band cloaking applications can be achieved if the hard surface (metasurface) concept \cite{Kildal,Kishk,Greenleaf} is employed to construct the cloaking device.

The proposed coordinate transformation technique \cite{Pendry,Leonhardt} was also used for the construction of elliptic \cite{Kwon} and square \cite{Rahm} cloaks. Moreover, it was applied to achieve cloaking in the acoustic frequency spectrum \cite{S.A.Cummer,Chan}. There are numerous novel applications which are based on the coordinate transformation method. The design of magnifying perfect and super lenses has been proposed in \cite{R.Smith,Psaltis}. A ``rotation coating'' which is derived with the same method of the implemented cloaking shell has been described in \cite{H.Chen}. A matched concentrator with free space in \cite{Rahm} and the design of conformal antennas in \cite{Y.Luo}. Furthermore, a reflectionless complex medium which can be utilized as an optical beam shifter or a beam splitter has been introduced in \cite{M.Rahm}.

Until now the lossless cloaking structure was modeled analytically \cite{Pendry,Leonhardt,D.Schurig}. A cylindrical wave expansion technique was occupied to simulate the lossless cloak in \cite{Ruan}. Recently an analytical method which is based on the Mie scattering model was proposed to exploit the lossy cloaking shell in the optical and microwave frequency region \cite{Chen}. The commercial simulation package COMSOL Multiphysics$^{\textmd{TM}}$ has been widely used to model different cloaks and to compare theoretical predictions \cite{Cummer,Cai,Chettiar,Kwon,Rahm}. It is based on a frequency domain numerical method, the finite element method (FEM). However, frequency domain techniques such as FEM can become inefficient if a wideband solution is desirable. Moreover, the cloak has been modeled analytically in the time-domain \cite{Weder} using a time-dependent scattering theory. The cloaking structure was firstly simulated with the finite-difference time-domain (FDTD) method in \cite{Yan}. Another FDTD cloaking modeling in which the Lorentz dispersive model is employed is presented in \cite{Liang}. In this paper, we propose a new radial-dependent dispersive FDTD method to model lossless and lossy cloaking devices and evaluate their bandwidth limitations. The auxiliary differential equation (ADE) method \cite{Gandhi} is used based on the Drude model to produce the FDTD updating equations. This dispersive FDTD method is a more general approach and an improvement of the previously proposed numerical technique in \cite{Yan}. The proposed method is able to fully exploit the cloaking phenomenon.

\section{Numerical Modeling of the Lossy Cloaking Structure}
The FDTD method is based on the temporal and spatial discretisation of Faraday's and Ampere's Laws which are:
\begin{eqnarray}
\nabla\times \bar{E}=-\frac{\partial \bar{B}}{\partial t}, \label{Faradaylaw}\\
\nabla\times \bar{H}=\frac{\partial \bar{D}}{\partial t} \label{Amperelaw}
\end{eqnarray}
where $\bar{E}$, $\bar{H}$, $\bar{D}$ and $\bar{B}$ are the electric field, magnetic field, electric flux density and magnetic flux density components respectively. However, for the dispersive FDTD method the constitutive equations have also to be discretised and they are given by the equations:
\begin{eqnarray}
\bar{D}=\varepsilon \bar{E}, \label{constitutiveelectricfield}\\
\bar{B}=\mu \bar{H} \label{constitutivemagneticfield}
\end{eqnarray}
where the permittivity $\varepsilon$ and permeability $\mu$ can have scalar or tensor form. For the following cloaking structure modeling the auxiliary differential equation (ADE) dispersive FDTD method will be employed. Faraday's and Ampere's Laws are discretised with the common procedure \cite{Taflove} and the conventional FDTD updating equations are:
\begin{eqnarray}
H^{n+1} = H^{n}-\left(\frac{\Delta t}{\mu}\right)\cdot\widetilde{\nabla}\times E^{n+\frac{1}{2}}, \label{discreteFaradaylaw}\\
E^{n+1} = E^{n}+\left(\frac{\Delta t}{\varepsilon}\right)\cdot\widetilde{\nabla}\times H^{n+\frac{1}{2}} \label{discreteAmperelaw}
\end{eqnarray}
where $\Delta t$ is the temporal discretisation of the FDTD method, $\widetilde{\nabla}$ is the discrete curl operator and $n$ the number of the current time step.

The full set of electromagnetic characteristic parameters of the cloaking structure in cylindrical coordinates is usually given as the following \cite{Cummer}:
\begin{eqnarray}
\varepsilon_{r}=\mu_{r}=\frac{r-R_{1}}{r},\hspace{2 mm}\varepsilon_{\phi}=\mu_{\phi}=\frac{r}{r-R_{1}},\hspace{2 mm}
\varepsilon_{z}=\mu_{z}=\left(\frac{R_{2}}{R_{2}-R_{1}}\right)^{2}\frac{r-R_{1}}{r}
\label{parametersofcloaking}
\end{eqnarray}
where $R_{1}$ is the inner radius, $R_{2}$ the outer radius and $r$ an arbitrary radius of the cloaking structure. From the equations (\ref{parametersofcloaking}) the ranges of the cloaking parameters are derived to be:
\begin{eqnarray*}
\varepsilon_{r},\mu_{r}\in\left[0,\frac{(R_{2}-R_{1})}{R_{1}}\right],\hspace{2 mm}\varepsilon_{\phi},\mu_{\phi}\in\left[\frac{R_{2}}{(R_{2}-R_{1})},\infty\right],\hspace{2 mm} \varepsilon_{z},\mu_{z}\in\left[0,\frac{R_{2}}{(R_{2}-R_{1})}\right].
\end{eqnarray*}

It is observed that the values of $\varepsilon_{r},\mu_{r},\varepsilon_{z},\mu_{z}$ are less than one for some points of $r$ and the values of $\varepsilon_{\phi},\mu_{\phi}$ are always bigger than one like conventional dielectrics. Thus, the conventional FDTD method cannot correctly simulate materials with the properties of $\varepsilon_{r},\mu_{r},\varepsilon_{z},\mu_{z}$ and new dispersive FDTD techniques have to be employed, like in the case of the left-handed metamaterials (LHMs) \cite{Belov}. The parameters are mapped with dispersive material models and more precisely with the well-known and widely used Drude model:
\begin{equation}
\varepsilon_{r}=1-\frac{\omega_{p}^2}{\omega^2-\jmath\omega\gamma}
\label{Drudemodel}
\end{equation}
where $\omega_{p}$ is the plasma frequency and $\gamma$ is the collision frequency which characterizes the losses of the dispersive material. The plasma frequency $\omega_{p}$ will vary in order to simulate the material properties of the radius dependent parameters (\ref{parametersofcloaking}). The required lossy permittivity can also be alternative presented by the formula $\hat{\varepsilon}_{r}=\varepsilon_{r}(1-\jmath\tan\delta)$, where $\varepsilon_{r}$ is radius dependent (\ref{parametersofcloaking}) and $\tan\delta$ is the loss tangent of the cloaking material. This formula is substituted in equation (\ref{Drudemodel}) and after straightforward derivations the plasma and collision frequencies can be calculated from the following analytical equations:
\begin{equation}
\omega_{p}^2=(1-\varepsilon_{r})\omega^2+\varepsilon_{r}\tan\delta\omega\gamma
\label{plasmafrequencyanalytical}
\end{equation}
\begin{equation}
\gamma=\frac{\varepsilon_{r}\tan\delta\omega}{(1-\varepsilon_{r})}
\label{collisionfrequencyanalytical}
\end{equation}
From the equations (\ref{plasmafrequencyanalytical}, \ref{collisionfrequencyanalytical}), it is obvious that both plasma and collision frequencies vary according to the radius of the cloaking device. Moreover, the plasma frequency is also dependent on the losses $\tan\delta, \gamma$ of the material.

The $\varepsilon_{\phi}$ parameter will always have values greater than one and it is simulated with the conventional lossy dielectric material model:
\begin{equation}
\hat{\varepsilon}_{\phi}=\varepsilon_{\phi}+\frac{\sigma}{\jmath\omega}
\label{lossydielectric}
\end{equation}
where the parameter $\varepsilon_{\phi}$ is dependent on the radius of cloaking shell as in equation (\ref{parametersofcloaking}) and $\sigma$ is a measurement of the conductivity losses. The loss tangent for the lossy dielectric material is given by $\tan\delta=\frac{\sigma}{\omega\varepsilon_{\phi}}$ and it is also radius dependent because it is a function of the $\varepsilon_{\phi}$ (\ref{parametersofcloaking}). The two-dimensional (2-D) transverse electric (TE) incidence is used during the simulations without lost of generality, which reduces the non-zero field to three components $E_{x},E_{y}$ and $H_{z}$. For TE wave polarisation only three parameters from the full set (\ref{parametersofcloaking}) are employed, the $\varepsilon_{r}$, $\varepsilon_{\phi}$ and $\mu_{z}$.

The classical Cartesian FDTD mesh is used in the modeling and the previously mentioned parameters are transformed from the cylindrical coordinates ($r,\phi,z$) to Cartesian ones ($x,y,z$) following the formulas given below:
\begin{eqnarray}
\varepsilon_{xx}=\varepsilon_{r}\cos^{2}{\phi}+\hat{\varepsilon}_{\phi}\sin^{2}{\phi},\hspace{2 mm} \varepsilon_{xy}=\varepsilon_{yx}=(\varepsilon_{r}-\hat{\varepsilon}_{\phi})\sin{\phi}\cos{\phi},\hspace{2 mm}
\varepsilon_{yy}=\varepsilon_{r}\sin^{2}{\phi}+\hat{\varepsilon}_{\phi}\cos^{2}{\phi}
\label{cartesianfromcylindrical}
\end{eqnarray}
Hence, the constitutive equation (\ref{constitutiveelectricfield}) written in a tensor form is given by:
\begin{eqnarray}
\left(\begin{array}{l}
D_{x}\\
D_{y} \end{array}\right)=\varepsilon_{0}\left(\begin{array}{ll}
\varepsilon_{xx} & \varepsilon_{xy}\\
\varepsilon_{yx} & \varepsilon_{yy}
\end{array}\right)
\left(\begin{array}{l}
E_{x}\\
E_{y} \end{array}\right)
\label{tensorformofconstitutiveequation}
\end{eqnarray}
From the equation (\ref{tensorformofconstitutiveequation}), it can be detected that:
\begin{eqnarray}
\left\{\begin{array}{l}
\varepsilon_{0}\varepsilon_{xx}E_{x}+\varepsilon_{0}\varepsilon_{xy}E_{y}=D_{x}\\
\varepsilon_{0}\varepsilon_{yx}E_{x}+\varepsilon_{0}\varepsilon_{yy}E_{y}=D_{y}
\end{array}
\right. \label{constitutiveequations}
\end{eqnarray}
where $\varepsilon_{xx},\varepsilon_{xy},\varepsilon_{yx},\varepsilon_{yy}$ are given in (\ref{cartesianfromcylindrical}). Substituting the $\varepsilon_{r}$ from the Drude model (\ref{Drudemodel}) and the lossy dielectric $\hat{\varepsilon}_{\phi}$ with the formula (\ref{lossydielectric}) in the first equation of (\ref{constitutiveequations}), it can be obtained in a straightforward manner:
\begin{eqnarray}
\varepsilon_{0}[\jmath\omega(\omega^{2}-\jmath\omega\gamma-\omega_{p}^2)\cos^{2}{\phi}+(\jmath\omega\varepsilon_{\phi}+\sigma)(\omega^{2}-\jmath\omega\gamma)\sin^{2}{\phi}]E_{x}\nonumber\\
+\varepsilon_{0}[\jmath\omega(\omega^{2}-\jmath\omega\gamma-\omega_{p}^2)-(\jmath\omega\varepsilon_{\phi}+\sigma)(\omega^{2}-\jmath\omega\gamma)]\sin{\phi}\cos{\phi}E_{y}=\jmath\omega(\omega^{2}-\jmath\omega\gamma)D_{x}
\label{constitutiveequationsbeforeFDTD}
\end{eqnarray}
Next, the above equation (\ref{constitutiveequationsbeforeFDTD}) is divided by $\jmath\omega$ to achieve a simpler lower order FDTD algorithm and it is transformed to:
\begin{eqnarray}
\varepsilon_{0}[(\omega^{2}-\jmath\omega\gamma-\omega_{p}^2)\cos^{2}{\phi}+(\varepsilon_{\phi}\omega^{2}-\jmath\omega(\sigma+\varepsilon_{\phi}\gamma)-\sigma\gamma)\sin^{2}{\phi}]E_{x}\nonumber\\
+\varepsilon_{0}[(\omega^{2}-\jmath\omega\gamma-\omega_{p}^2)-(\varepsilon_{\phi}\omega^{2}-\jmath\omega(\sigma+\varepsilon_{\phi}\gamma)-\sigma\gamma)]\sin{\phi}\cos{\phi}E_{y}=(\omega^{2}-\jmath\omega\gamma)D_{x}
\label{constitutiveequationsbeforeFDTDand2order}
\end{eqnarray}

Thus, the updating dispersive FDTD equation can be obtained from the formula (\ref{constitutiveequationsbeforeFDTDand2order}) via the inverse Fourier transform ($\jmath\omega\rightarrow\frac{\partial}{\partial t}$, $\omega^{2}\rightarrow-\frac{\partial^{2}}{\partial t^{2}}$) and the equation (\ref{constitutiveequationsbeforeFDTDand2order}) becomes:
\begin{eqnarray}
\varepsilon_{0}[(\frac{\partial^{2}}{\partial t^{2}}+\gamma\frac{\partial}{\partial t}+\omega_{p}^2)\cos^{2}{\phi}+(\varepsilon_{\phi}\frac{\partial^{2}}{\partial t^{2}}+(\sigma+\varepsilon_{\phi}\gamma)\frac{\partial}{\partial t}+\sigma\gamma)\sin^{2}{\phi}]E_{x}\nonumber\\
+\varepsilon_{0}[(\frac{\partial^{2}}{\partial t^{2}}+\gamma\frac{\partial}{\partial t}+\omega_{p}^2)-(\varepsilon_{\phi}\frac{\partial^{2}}{\partial t^{2}}+(\sigma+\varepsilon_{\phi}\gamma)\frac{\partial}{\partial t}+\sigma\gamma)]\sin{\phi}\cos{\phi}E_{y}=(\frac{\partial^{2}}{\partial t^{2}}+\gamma\frac{\partial}{\partial t})D_{x}
\label{constitutiveequationsafterfouriertrans}
\end{eqnarray}
A second-order discretisation procedure is applied in equation (\ref{constitutiveequationsafterfouriertrans}) where the central finite difference operators in time ($\delta_{t}$ and $\delta_{t}^{2}$) and the central average operators with respect to time ($\mu_{t}$ and $\mu_{t}^2$) are used:
\begin{equation}
\frac{\partial^{2}}{\partial t^{2}}\rightarrow\frac{\delta_{t}^{2}}{\Delta t^2},\hspace{2 mm}\frac{\partial}{\partial t}\rightarrow\frac{\delta_{t}}{\Delta t}\mu_{t}\hspace{2 mm},\omega_{p}^2\rightarrow\omega_{p}^2\mu_{t}^2,\hspace{2 mm}\sigma\gamma\rightarrow\sigma\gamma\mu_{t}^2\label{operators}
\end{equation}
where the operators $\delta_{t},\delta_{t}^{2},\mu_{t},\mu_{t}^2$ are explained in \cite{Hildebrand} and they are given by:
\begin{eqnarray}
\delta_{t}F|_{i,j,k}^{n}\equiv F|_{i,j,k}^{n+\frac{1}{2}}-F|_{i,j,k}^{n-\frac{1}{2}},\hspace{2 mm}
\delta_{t}^{2}F|_{i,j,k}^{n}\equiv F|_{i,j,k}^{n+1}-2F|_{i,j,k}^{n}+F|_{i,j,k}^{n-1},\nonumber\\
\mu_{t}F|_{i,j,k}^{n}\equiv \frac{F|_{i,j,k}^{n+\frac{1}{2}}+F|_{i,j,k}^{n-\frac{1}{2}}}{2},\hspace{2 mm}
\mu_{t}^2F|_{i,j,k}^{n}\equiv \frac{F|_{i,j,k}^{n+1}+2F|_{i,j,k}^{n}+F|_{i,j,k}^{n-1}}{4}\label{operatorsexplained}
\end{eqnarray}
The \textit{F} represents arbitrary field components and the $(i,j,k)$ indices are the coordinates of a certain mesh point in the FDTD domain. Hence, the discretised equation (\ref{constitutiveequationsafterfouriertrans}) is the following:
\begin{eqnarray}
\varepsilon_{0}[(\frac{\delta_{t}^{2}}{\Delta t^2}+\gamma\frac{\delta_{t}}{\Delta t}+\omega_{p}^2\mu_{t}^2)\cos^{2}{\phi}+(\varepsilon_{\phi}\frac{\delta_{t}^{2}}{\Delta t^2}+(\sigma+\varepsilon_{\phi}\gamma)\frac{\delta_{t}}{\Delta t}+\sigma\gamma\mu_{t}^2)\sin^{2}{\phi}]E_{x}\nonumber\\
+\varepsilon_{0}[(\frac{\delta_{t}^{2}}{\Delta t^2}+\gamma\frac{\delta_{t}}{\Delta t}+\omega_{p}^2\mu_{t}^2)-(\varepsilon_{\phi}\frac{\delta_{t}^{2}}{\Delta t^2}+(\sigma+\varepsilon_{\phi}\gamma)\frac{\delta_{t}}{\Delta t}+\sigma\gamma\mu_{t}^2)]\sin{\phi}\cos{\phi}E_{y}=(\frac{\delta_{t}^{2}}{\Delta t^2}+\gamma\frac{\delta_{t}}{\Delta t})D_{x}
\label{constitutiveequationsafterdiscretisation}
\end{eqnarray}

Note that the $\varepsilon_{\phi}$ parameter remains constant in equation (\ref{constitutiveequationsafterdiscretisation}) because it is always greater than one as the conventional dielectric materials. Finally, the operators (\ref{operatorsexplained}) are substituted in equation (\ref{constitutiveequationsafterdiscretisation}) and the derived dispersive FDTD updating equation is shown in the following:
\begin{eqnarray}
E_{x}^{n+1}=[C_{1}D_{x}^{n+1}-B_{1}E_{y}^{n+1}-C_{2}{D}_{x}^{n}+A_{2}E_{x}^{n}+B_{2}E_{y}^{n}+C_{3}D_{x}^{n-1}-A_{3}E_{x}^{n-1}-B_{3}E_{y}^{n-1}]/A_{1}
\label{FDTDExequation}
\end{eqnarray}
With exactly the same method, the FDTD updating equation for the $E_{y}$ component can be derived from the second equation of (\ref{constitutiveequations}) as:
\begin{eqnarray}
E_{y}^{n+1}=[C_{1}D_{y}^{n+1}-B_{1}E_{x}^{n+1}-C_{2}{D}_{y}^{n}+F_{2}E_{y}^{n}+B_{2}E_{x}^{n}+C_{3}D_{y}^{n-1}-F_{3}E_{y}^{n-1}-B_{3}E_{x}^{n-1}]/F_{1}
\label{FDTDEyequation}
\end{eqnarray}
The coefficients for both equations (\ref{FDTDExequation}), (\ref{FDTDEyequation}) are given by:
\begin{eqnarray*}
A_{1}=\frac{(\cos^{2}{\phi}+\varepsilon_{\phi}\sin^{2}{\phi})}{\Delta t^2}+\frac{\omega_{p}^2\cos^{2}{\phi}+\sigma\gamma\sin^{2}{\phi}}{4}+\frac{\gamma\cos^{2}{\phi}+(\sigma+\varepsilon_{\phi}\gamma)\sin^{2}{\phi}}{2\Delta t},\\
A_{2}=\frac{2(\cos^{2}{\phi}+\varepsilon_{\phi}\sin^{2}{\phi})}{\Delta t^2}-\frac{\omega_{p}^2\cos^{2}{\phi}+\sigma\gamma\sin^{2}{\phi}}{2},\\
A_{3}=\frac{(\cos^{2}{\phi}+\varepsilon_{\phi}\sin^{2}{\phi})}{\Delta t^2}+\frac{\omega_{p}^2\cos^{2}{\phi}+\sigma\gamma\sin^{2}{\phi}}{4}-\frac{\gamma\cos^{2}{\phi}+(\sigma+\varepsilon_{\phi}\gamma)\sin^{2}{\phi}}{2\Delta t},\\
B_{1}=\frac{(1-\varepsilon_{\phi})\sin{\phi}\cos{\phi}}{\Delta t^2}+\frac{(\omega_{p}^2-\sigma\gamma)\sin{\phi}\cos{\phi}}{4}+\frac{(\gamma-\sigma-\varepsilon_{\phi}\gamma)\sin{\phi}\cos{\phi}}{2\Delta t},\\
B_{2}=\frac{2(1-\varepsilon_{\phi})\sin{\phi}\cos{\phi}}{\Delta t^2}-\frac{(\omega_{p}^2-\sigma\gamma)\sin{\phi}\cos{\phi}}{2},\\
B_{3}=\frac{(1-\varepsilon_{\phi})\sin{\phi}\cos{\phi}}{\Delta t^2}+\frac{(\omega_{p}^2-\sigma\gamma)\sin{\phi}\cos{\phi}}{4}-\frac{(\gamma-\sigma-\varepsilon_{\phi}\gamma)\sin{\phi}\cos{\phi}}{2\Delta t},\\
C_{1}=\frac{1}{\varepsilon_{0}\Delta t^2}+\frac{\gamma}{2\varepsilon_{0}\Delta t}, C_{2}=\frac{2}{\varepsilon_{0}\Delta t^2}, C_{3}=\frac{1}{\varepsilon_{0}\Delta t^2}-\frac{\gamma}{2\varepsilon_{0}\Delta t},\\
F_{1}=\frac{(\sin^{2}{\phi}+\varepsilon_{\phi}\cos^{2}{\phi})}{\Delta t^2}+\frac{\omega_{p}^2\sin^{2}{\phi}+\sigma\gamma\cos^{2}{\phi}}{4}+\frac{\gamma\sin^{2}{\phi}+(\sigma+\varepsilon_{\phi}\gamma)\cos^{2}{\phi}}{2\Delta t},\\
F_{2}=\frac{2(\sin^{2}{\phi}+\varepsilon_{\phi}\cos^{2}{\phi})}{\Delta t^2}-\frac{\omega_{p}^2\sin^{2}{\phi}+\sigma\gamma\cos^{2}{\phi}}{2},\\
F_{3}=\frac{(\sin^{2}{\phi}+\varepsilon_{\phi}\cos^{2}{\phi})}{\Delta t^2}+\frac{\omega_{p}^2\sin^{2}{\phi}+\sigma\gamma\cos^{2}{\phi}}{4}-\frac{\gamma\sin^{2}{\phi}+(\sigma+\varepsilon_{\phi}\gamma)\cos^{2}{\phi}}{2\Delta t}.
\end{eqnarray*}
where $\Delta t$ is the temporal discretisation of the FDTD method.

However, the above equations (\ref{FDTDExequation}), (\ref{FDTDEyequation}) cannot be calculated with the FDTD algorithm. The reason is that in the case (\ref{FDTDExequation}) the component $E_{y}^{n+1}$ cannot be computed at the particular time step ($n+1$). This also applies to $E_{x}^{n+1}$ component in the equation (\ref{FDTDEyequation}). The solution is to substitute equation (\ref{FDTDEyequation}) into the formula (\ref{FDTDExequation}) and the inverse. As a result, the updating FDTD equation which computes the $E_{x}^{n+1}$ component becomes:
\begin{eqnarray}
E_{x}^{n+1}=[C_{1}D_{x}^{n+1}-a_{1}\overline{D}_{y}^{n+1}-C_{2}{D}_{x}^{n}+a_{2}\overline{D}_{y}^{n}+b_{2}E_{x}^{n}+d_{1}\overline{E}_{y}^{n}+C_{3}D_{x}^{n-1}-a_{3}\overline{D}_{y}^{n-1}-b_{3}E_{x}^{n-1}\nonumber\\
-d_{2}\overline{E}_{y}^{n-1}]/b_{1}
\label{FDTDequationfinal}
\end{eqnarray}
The updating FDTD equation for the $E_{y}^{n+1}$ component is found with exactly the same way to be:
\begin{eqnarray}
E_{y}^{n+1}=[C_{1}D_{y}^{n+1}-e_{1}\overline{D}_{x}^{n+1}-C_{2}{D}_{y}^{n}+e_{2}\overline{D}_{x}^{n}+f_{2}E_{y}^{n}+g_{1}\overline{E}_{x}^{n}+C_{3}D_{y}^{n-1}-e_{3}\overline{D}_{x}^{n-1}-f_{3}E_{y}^{n-1}\nonumber\\
-g_{2}\overline{E}_{x}^{n-1}]/f_{1}
\label{FDTDequationfinalEy}
\end{eqnarray}
where the newly introduced coefficients in equations (\ref{FDTDequationfinal}) and (\ref{FDTDequationfinalEy}) are:
\begin{eqnarray*}
a_{1}=\frac{B_{1}C_{1}}{F_{1}}, a_{2}=\frac{B_{1}C_{2}}{F_{1}}, a_{3}=\frac{B_{1}C_{3}}{F_{1}},
b_{1}=A_{1}-\frac{B_{1}^2}{F_{1}}, b_{2}=A_{2}-\frac{B_{1}B_{2}}{F_{1}}, b_{3}=A_{3}-\frac{B_{1}B_{3}}{F_{1}},\\
d_{1}=B_{2}-\frac{B_{1}F_{2}}{F_{1}}, d_{2}=B_{3}-\frac{B_{1}F_{3}}{F_{1}},
e_{1}=\frac{B_{1}C_{1}}{A_{1}}, e_{2}=\frac{B_{1}C_{2}}{A_{1}}, e_{3}=\frac{B_{1}C_{3}}{A_{1}},\\
f_{1}=F_{1}-\frac{B_{1}^2}{A_{1}}, f_{2}=F_{2}-\frac{B_{1}B_{2}}{A_{1}}, f_{3}=F_{3}-\frac{B_{1}B_{3}}{A_{1}},
g_{1}=B_{2}-\frac{A_{2}B_{1}}{A_{1}}, g_{2}=B_{3}-\frac{A_{3}B_{1}}{A_{1}}.
\end{eqnarray*}
For more accurate results, the overlined field components $\overline{D}_{y},\overline{E}_{y},\overline{D}_{x},\overline{E}_{x}$ are calculated with a locally spatial averaging technique \cite{Lee}. This method is employed because the $x$ and $y$ field components are located in different mesh points across the FDTD grid. Their averaged values are computed from the next formula \cite{Lee}:
\begin{eqnarray}
\overline{E}_{y}(i,j)=[E_{y}(i,j)+E_{y}(i+1,j)+E_{y}(i,j-1)+E_{y}(i+1,j-1)]/4
\label{spatialaverage}
\end{eqnarray}
where $(i,j)$ the coordinates of the mesh point.

The final step is to introduce the updating FDTD equation of the $H_{z}$ field component. From the equation (\ref{parametersofcloaking}), the magnetic permeability $\mu_{z}$ component can have values less and bigger than one. Hence a more complicated approach is necessary to model the magnetic field $H_{z}$ component. When $\mu_{z}<1$, the magnetic permeability is mapped with the Drude model and it is given from the formula:
\begin{equation}
\mu_{z}=1-\frac{\omega_{pm}^2}{\omega^2-\jmath\omega\gamma_{m}}
\label{Drudemodelmagnetic}
\end{equation}
where $\omega_{pm}$ is the magnetic plasma frequency and $\gamma_{m}$ is the magnetic collision frequency which measures the losses of the magnetic dispersive material. The analytical equations of $\omega_{pm}$ and $\gamma_{m}$ are derived with the same way of equations (\ref{plasmafrequencyanalytical}), (\ref{collisionfrequencyanalytical}) and they are given by:
\begin{equation}
\omega_{pm}^2=(1-\mu_{z})\omega^2+\mu_{z}\tan\delta_{m}\omega\gamma_{m}
\label{magneticplasmafrequencyanalytical}
\end{equation}
\begin{equation}
\gamma_{m}=\frac{\mu_{z}\tan\delta_{m}\omega}{(1-\mu_{z})}
\label{magneticcollisionfrequencyanalytical}
\end{equation}
From the above equations (\ref{magneticplasmafrequencyanalytical}), (\ref{magneticcollisionfrequencyanalytical}) it is derived that the magnetic plasma and collision frequencies are radius dependent because they are functions of $\mu_{z}$ (\ref{parametersofcloaking}).

The formula (\ref{Drudemodelmagnetic}) is substituted in the constitutive equation (\ref{constitutivemagneticfield}) and it is discretised as in \cite{Belov}. The FDTD updating equation for this case is:
\begin{eqnarray}
H_{z}^{n+1}=\left\{\left[\frac{1}{\mu_{0}\Delta t^2}+\frac{\gamma_{m}}{2\mu_{0}\Delta t}\right]B_{z}^{n+1}-\frac{2}{\mu_{0}\Delta t^2}B_{z}^{n}+\left[\frac{1}{\mu_{0}\Delta t^2}-\frac{\gamma_{m}}{2\mu_{0}\Delta t}\right]B_{z}^{n-1}+\left[\frac{2}{\Delta t^2}-\frac{\omega_{pm}^2}{2}\right]H_{z}^{n}\right.\nonumber\\
\left.-\left[\frac{1}{\Delta t^2}-\frac{\gamma_{m}}{2\Delta t}+\frac{\omega_{pm}^2}{4}\right]H_{z}^{n-1}\right\}\Bigg{/}\left[\frac{1}{\Delta t^2}-\frac{\gamma_{m}}{2\Delta t}+\frac{\omega_{pm}^2}{4}\right]
\label{FDTDequationfinalHz}
\end{eqnarray}
When the magnetic permeability of the cloaking material is $\mu_{z}\geq1$, it is simulated with the conventional lossy magnetic model:
\begin{equation}
\hat{\mu}_{z}=\mu_{z}+\frac{\sigma_{m}}{\jmath\omega}
\label{lossymagnetic}
\end{equation}
where the component $\mu_{z}$ is radius dependent and it is given from the equation (\ref{parametersofcloaking}). The parameter $\sigma_{m}$ is the magnetic conductivity. The loss tangent for the lossy magnetic material is equal to $\tan\delta_{m}=\frac{\sigma_{m}}{\omega\mu_{z}}$ and it is also radius dependent. The updating FDTD equation for this type of material is derived from the discrete Faraday law (\ref{discreteFaradaylaw}) with losses included \cite{Taflove}. Finally, the FDTD updating equation of $H$ to $E$ is equal to the discrete Ampere law (\ref{discreteAmperelaw}) in free space.

Note that the method in \cite{Yan} for a full set of cloaking parameters (eq. (\ref{parametersofcloaking})) cannot be applied when $\varepsilon_{r}=0$ which occurs only for the points at the inner radius of the cloaking device ($r=R_{1}$). More precisely, the permittivity tensor $\{\varepsilon\}$ (\ref{tensorformofconstitutiveequation}) cannot be inverted when $\varepsilon_{r}=0$ because its determinant is equal to: $|\varepsilon|=\varepsilon_{r}\varepsilon_{\phi}$. From the equations (\ref{parametersofcloaking}), it is obtained that $\varepsilon_{\phi}\neq0$ for all the cases. Thus, if $\varepsilon_{r}=0$ the determinant $|\varepsilon|$ is zero and the array $\{\varepsilon\}$ does not have an inverse. Furthermore, for the method presented in \cite{Yan}, the conductive losses of the $\hat{\varepsilon}_{\phi}$ component and the conventional magnetic losses of the $\hat{\mu}_{z}$ parameter are not introduced in the updating FDTD equations. Therefore, the currently proposed FDTD method is an extension of the one proposed in \cite{Yan} and it can also simulate lossy electromagnetic cloaks. The proposed method is ready to be extended to model three dimensional (3-D) lossy electromagnetic cloaks.

Numerical approximations are inevitable when the FDTD method is applied. Space and time are discretised with a detrimental effect on the accuracy of the simulations. Furthermore, the permittivity and permeability are frequency dependent as the Drude model (\ref{Drudemodel}) and the lossy dielectric behaviour (\ref{lossydielectric}) are used. Due to the presence of a discrete time step $\Delta t$ used in the FDTD method there will be differences between the analytical and the numerical characteristics of the cloaking material. Hence, for the proposed dispersive FDTD method a spatial resolution of $\Delta x<\lambda/10$ is not sufficient unlike the conventional dielectric material simulations in which it is the required value \cite{Taflove}. From a previous analysis of left-handed metamaterials (LHMs) \cite{Belov}, it was found that spurious resonances are caused due to coarse time discretisation which leads to numerical errors and inaccurate modeling results. It was proposed that a spatial resolution of $\Delta x<\lambda/80$ is essential for accurate simulations. The same and more dense spatial resolution restrictions have to be applied in the simulation of the cloaking structure.

The same approach will be followed with \cite{Belov,Yan} in order to compute the numerical values of the permittivities $\varepsilon_{r},\varepsilon_{\phi}$ and the permeability $\mu_{z}$. The plane waves in a discrete time form are:
\begin{equation}
E^{n}=Ee^{\jmath n\omega\Delta t},D^{n}=De^{\jmath n\omega\Delta t}
\label{planewaves}
\end{equation}
They are substituted in equation (\ref{FDTDExequation}) and the calculated numerical permittivities $\widetilde{\varepsilon_{r}},\widetilde{\varepsilon_{\phi}}$ are becoming:
\begin{equation}
\widetilde{\varepsilon_{r}}=\left[1-\frac{\omega_{p}^2\Delta t^2\cos^{2}{\frac{\omega\Delta t}{2}}}{2\sin{\frac{\omega\Delta t}{2}}(2\sin{\frac{\omega\Delta t}{2}}-\jmath\gamma\Delta t\cos{\frac{\omega\Delta t}{2}})}\right]
\label{Drudemodelcorrected}
\end{equation}
\begin{equation}
\widetilde{\varepsilon_{\phi}}=\varepsilon_{\phi}+\frac{\sigma\Delta t}{2\jmath \tan{\frac{\omega\Delta t}{2}}}
\label{lossydielectriccorrected}
\end{equation}
Notice that when $\Delta t \rightarrow 0$ which leads to a very fine FDTD grid, the equations (\ref{Drudemodelcorrected},\ref{lossydielectriccorrected}) are transformed to the Drude model (\ref{Drudemodel}) and the lossy dielectric material (\ref{lossydielectric}) respectively. The exact same numerical permeability $\widetilde{\mu_{z}}$ formulas can be produced for the dispersive magnetic model (\ref{Drudemodelmagnetic}) and the conventional lossy magnetic material (\ref{lossymagnetic}). The comparison between analytic and numerical material parameters can be seen in \cite{Yan}. It can be concluded that the conventional spatial resolutions $\Delta x<\lambda/10$ are not appropriate for this kind of anisotropic materials and more fine FDTD meshes with $\Delta x<\lambda/80$ have to be used for more accurate simulations.

Another problem which was dominant during the FDTD modeling of cloaking structure is numerical instability. The Courant stability criterion $\Delta t =\Delta x/\sqrt{2}c$ \cite{Taflove} was satisfied during the FDTD simulations. The object which was ``cloaked'' was chosen to be composed of a perfect electric conductor (PEC) material. Arbitrary materials can be used for the object placed inside the cloaking shell. However for the FDTD modeling, it is better to choose PEC material because very small field values will always be expected inside the cloaked space due to the numerical approximations which are inherent to the FDTD method. The instability was generated at two specific regions of the cloaking FDTD meshes. The first instability region was obtained at the interface between the cloaking material and the free space ($r=R_{2}$). The other was concentrated at the interface between the cloaking device and the ``cloaked'' PEC material ($r=R_{1}$). In both regions the permittivities $\varepsilon_{r},\varepsilon_{\phi}$ and the permeability $\mu_{z}$ are changing rapidly from finite, even zero, to infinite theoretical values. As a result, spurious cavity resonances are created which are combined with the irregular staircase approximation of the cloaking structure's cylindrical geometry. From the discretisation with the FDTD method of the divergence of the electric flux density $\nabla\cdot D$, it can be concluded that the instability is present in the form of accumulated charges at the two interfaces.

In order to achieve stable FDTD simulations a series of modifications have to be applied in the conventional FDTD algorithm. Firstly, the locally spatial averaging technique (\ref{spatialaverage}) has to be introduced for the simulation of the constitutive equation which is in tensor form (\ref{tensorformofconstitutiveequation}). This improves the stability and accuracy of the cloaking modeling. Fine spatial resolutions ($\Delta x<\lambda/80$) have to be applied which will alleviate the effect of the inevitable for the current geometry staircase approximations. Ideally, an infinite spatial resolution will guarantee an accurate and stable cloaking modeling. Moreover, there are differences between the analytical and the numerical (\ref{Drudemodelcorrected}) material parameters which are affecting the stability of the FDTD simulations in a straightforward manner. Corrected numerical electric and magnetic plasma and collision frequencies have to be computed. The required numerical lossy permittivity is equal to $\widetilde{\varepsilon_{r}}=\varepsilon_{r}(1-\jmath\tan\delta)$, where $\varepsilon_{r}$ is radial-dependent (\ref{parametersofcloaking}) and $\tan\delta$ is the loss tangent of the cloaking material. If the numerical lossy permittivity is substituted in the equation (\ref{Drudemodelcorrected}) the resulted corrected plasma and collision frequencies are obtained as the following \cite{Yan}:
\begin{equation}
\widetilde{\omega}_{p}^2=\frac{2\sin{\frac{\omega\Delta t}{2}}[-2(\varepsilon_{r}-1)\sin{\frac{\omega\Delta t}{2}}+\varepsilon_{r}\tan{\delta}\gamma\Delta t \cos{\frac{\omega\Delta t}{2}}]}{\Delta t^2\cos^{2}{\frac{\omega\Delta t}{2}}}
\label{plasmafrequencycorrected}
\end{equation}
\begin{equation}
\widetilde{\gamma}=\frac{2\varepsilon_{r}\tan{\delta}\sin{\frac{\omega\Delta t}{2}}}{(1-\varepsilon_{r})\Delta t \cos{\frac{\omega\Delta t}{2}}}
\label{collisionfrequencycorrected}
\end{equation}
For the conventional lossy dielectric or magnetic model the only correction for improved stability is for the frequency as it can be derived from equation (\ref{lossydielectriccorrected}):
\begin{equation}
\widetilde{\omega}=\frac{\tan{\frac{\omega\Delta t}{2}}}{\Delta t/2}
\label{frequencycorrected}
\end{equation}

With all the previous modifications a stable FDTD simulation can be satisfied at the outer interface ($r=R_{2}$) of the cloaking material. For the inner interface ($r=R_{1}$) one more modification has to be applied in the FDTD algorithm in order to achieve stability. The correct definition of the ``cloaked'' perfect electric conductor (PEC) material is crucial for stable modeling of the cloaking structure. The PEC is defined in the FDTD code as a material with infinite permittivity ($\varepsilon\rightarrow\infty$). Hence, the coefficient $\left(\frac{\Delta t}{\varepsilon}\right)$ in the discrete Ampere's Law (\ref{discreteAmperelaw}) has to be set to zero inside the PEC material for a correct and stable simulation. After all the previous mentioned modifications in the FDTD algorithm the resulted modeling is stable and the numerical accuracy has been improved. There are no accumulated charges at the two interfaces ($r=R_{1},r=R_{2}$) and this becomes evident by the FDTD simulation of the divergence of the electric flux density $\nabla\cdot D$.

\section{Numerical Results}
A 2-D TE polarisation case is used for the dispersive FDTD modeling of the cloaking device. A plane wave source is utilized to illuminate the cloaking structure. Uniform spatial discretisation is used, with the size of every FDTD cell to be $\Delta x=\Delta y=\lambda/150$, where $\lambda$ is the wavelength of excitation signals. In this case the operating frequency is $f=2$ GHz and the wavelength is $\lambda=15$ cm. The temporal discretisation is chosen according to the Courant stability condition \cite{Taflove} and the time step is given by $\Delta t=\Delta x/\sqrt{2}c$, where $c$ is the speed of light in free space.

\begin{figure}[h]
\centering
\includegraphics[width=8.0cm]{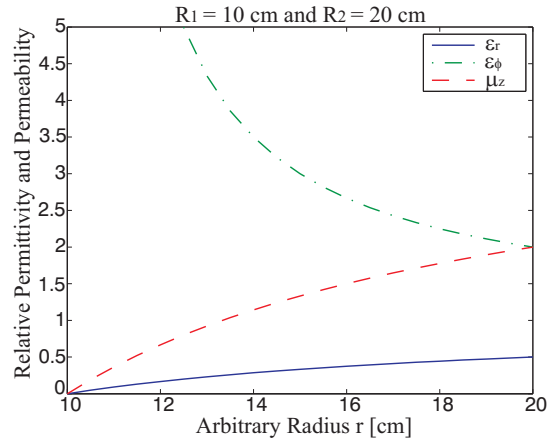}
\caption{Full set of cloaking material parameters used in the FDTD simulation.} \label{fullparameterset}
\end{figure}

\begin{figure}[h]
\centering
\includegraphics[width=8.0cm]{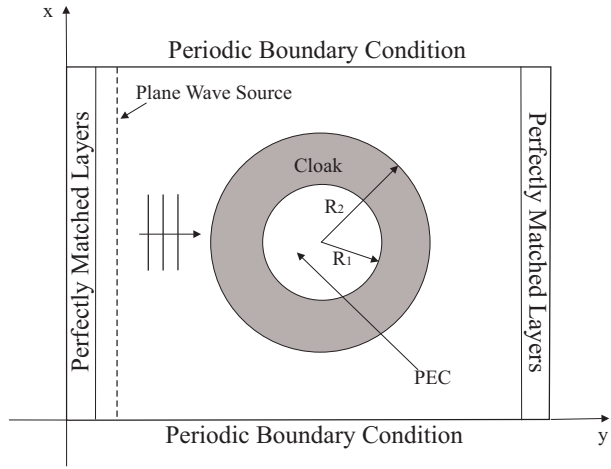}
\caption{A two-dimensional (2-D) FDTD computation domain of the cloaking structure for the case of plane wave excitation.} \label{fdtddomainofcloaking}
\end{figure}

\begin{figure}[h]
\centering
\includegraphics[width=9cm]{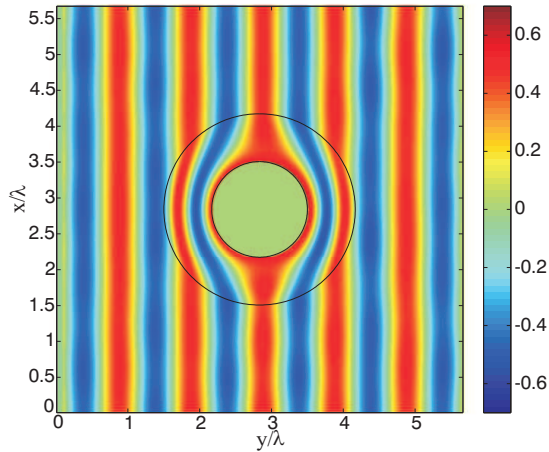}
\caption{Normalized magnetic field distribution of the lossless cloaking device with plane wave excitation. The wave propagates from left to right and the cloaked object is composed of a perfect electric conductor (PEC) material} \label{cloakingdeviceplanewave}
\end{figure}

\begin{figure}[h]
\centering
\includegraphics[width=8cm]{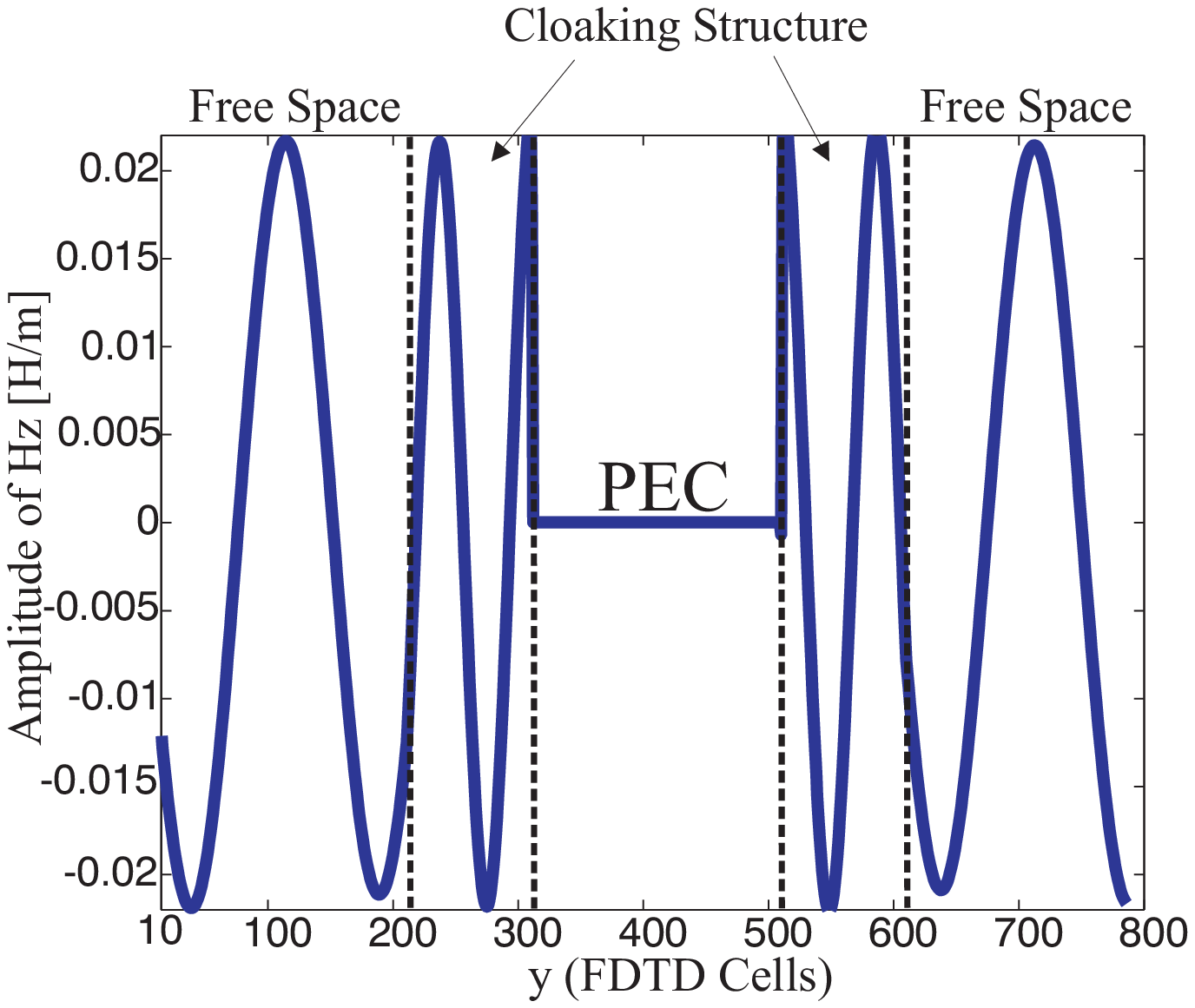}
\caption{A transverse profile of the magnetic field $H_{z}$ component propagating through the lossless cloaking device. The wave propagates from left to right undisturbed.} \label{cloakingdevicelosslesshzcomponent}
\end{figure}

Firstly, the lossless cloaking shell is simulated to validate the proposed FDTD method which means that the collision frequency in Drude model (\ref{Drudemodel}) is equal to zero $\gamma=0$. Furthermore, the conductivity in equation (\ref{lossydielectric}) is set to be zero $\sigma=0$. Hence, the radial dependent plasma frequency is computed from the equation $\omega_{p}=\omega\sqrt{1-\varepsilon_{r}}$ and the $\varepsilon_{r}$ is given by the equation (\ref{parametersofcloaking}). The inner and outer radius of the cloaking device have dimensions $R_{1}=10$ cm and $R_{2}=20$ cm respectively. The full set of the cloaking parameters (\ref{parametersofcloaking}) is changing as shown in the Fig. \ref{fullparameterset}.
The computational domain is terminated along the y-direction with Berenger's perfectly matched layer (PML) \cite{Berenger}. The waves are absorbed and it is the same as they are leaving from the computation domain without introducing reflections. In the last layer of the computational domain along the x-direction Bloch's periodic boundary conditions (PBCs) \cite{Taflove} are applied in order to create a propagating plane wave. The FDTD computation domain for this case can be obtained in the Fig. \ref{fdtddomainofcloaking}.
The results for plane wave excitation when the steady-state is reached can be seen in Fig. \ref{cloakingdeviceplanewave}. A transverse profile of the propagating field in the lossless cloaking shell is depicted in Fig. \ref{cloakingdevicelosslesshzcomponent}.

In Fig. \ref{cloakingdeviceplanewave} the electromagnetic wave is propagating from the left to the right side of the FDTD computation domain. The wave bends inside the cloaking device in order to avoid the ``cloaked'' object as expected. The wave trajectory is recomposed without any disturbance behind the cloaking shell. Therefore, the object inside the cloaking structure is like it does not exist, like it is ``invisible''. Mind that for this type of cloaking devices there are no constraints about the size and the material type of the object which will become ``invisible''. This is in contrast to the properties of the proposed left-handed metamaterials (LHMs) cloaking devices \cite{Alu,Milton,Abajo}. In Fig. \ref{cloakingdeviceplanewave}, small disturbance of the plane wave which is coming out of the cloaking shell on the right side can be seen. Furthermore, there is a slight scattering coming back to the source plane at the left side. The reason is that the surface of the cloaking device is curved (cylindrical structure) and it is being modeled with Cartesian FDTD mesh. As a result, staircase approximation is inevitable which directly reduces the simulation accuracy. This problem can be solved if a conformal scheme is utilized with a dispersive FDTD scheme \cite{Zhao} or a cylindrical FDTD is applied. However, the analysis of the conformal dispersive FDTD \cite{Zhao} for the cloaking structure leads to a complicated sixth-order differential equation. This is due to the anisotropy of the cloaking material parameters.

\begin{figure}[h]
\centering
\includegraphics[width=9cm]{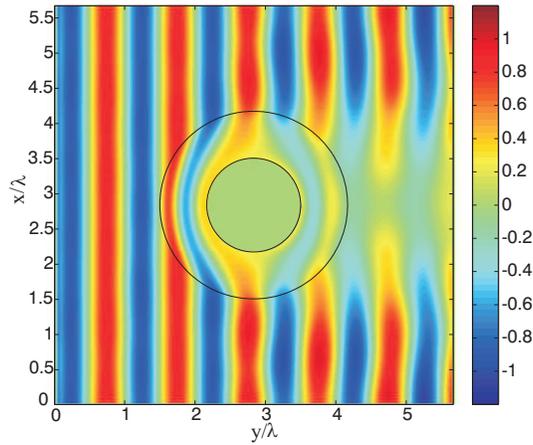}
\caption{Normalized magnetic field distribution of the lossy cloaking device with plane wave excitation. Ideal parameters are used with a loss tangent of ($\tan\delta=0.1$). The wave propagates from left to right and the cloaked object is composed of a perfect electric conductor (PEC) material.} \label{cloakingdevicelossy0.1}
\end{figure}

\begin{figure}[h]
\centering
\includegraphics[width=9cm]{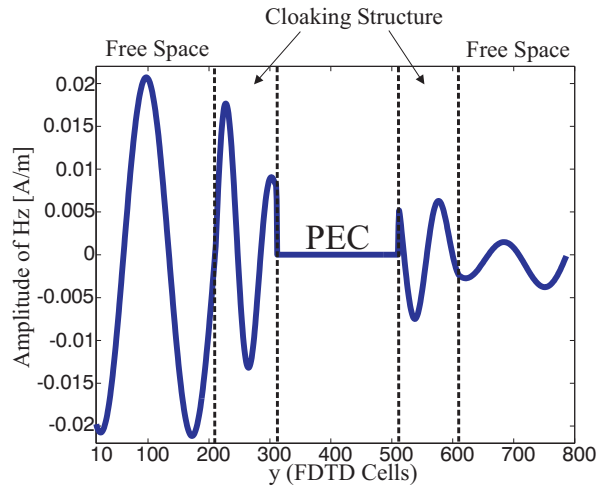}
\caption{Magnetic field $H_{z}$ component propagating through the lossy cloaking device ($\tan\delta=0.1$). The wave propagates from left to right and it is dissipated at the right side of the cloak.} \label{cloakingdevicelossyhzcomponent}
\end{figure}

The next step is to introduce losses in the radius dependent and dispersive cloaking material which is a far more practical and realistic representation of the metamaterials. The loss tangent is chosen equal to $\tan\delta=0.1$ for both the dispersive $\varepsilon_{r}$ component (\ref{Drudemodel}) and the conventional lossy dielectric component $\hat{\varepsilon}_{\phi}$ (\ref{lossydielectric}). For the magnetic component $\mu_{z}$, the magnetic loss tangent is chosen to be $\tan\delta_{m}=0.1$ for both the dispersive (\ref{Drudemodelmagnetic}) and the conventional lossy (\ref{lossymagnetic}) cases. The FDTD computational domain scenario which is used for simulating the lossy cloaking shell is again the same as Fig. (\ref{fdtddomainofcloaking}). The magnetic field $H_{z}$ distribution with a plane wave excitation is depicted in Fig. \ref{cloakingdevicelossy0.1}. It can be observed that the cloaking device is working (bending of waves) properly like the lossless case. But due to the presence of losses in electromagnetic cloaks, there is a shadowing effect to the field behind the cloaking shell. Nevertheless, it can be seen that the backscatter performance is improved in comparison with the lossless case as was mentioned in \cite{Cummer}. The attenuation of the propagating magnetic field $H_{z}$ component through the lossy cloaking shell can be seen also in Fig. \ref{cloakingdevicelossyhzcomponent}. For $\tan\delta=0.01$ the magnetic field pattern is almost identical to the ideal lossless case in Fig. \ref{cloakingdeviceplanewave}. However, the cloaking performance is impaired due to the casted shadow behind the cloaked object for the case of loss tangent $\tan\delta=0.1$. Therefore, the proposed cloaking structure is sensitive to losses which is a drawback towards the realization of future ``invisibility'' devices.

\begin{figure}[h]
\centering
\includegraphics[width=9cm]{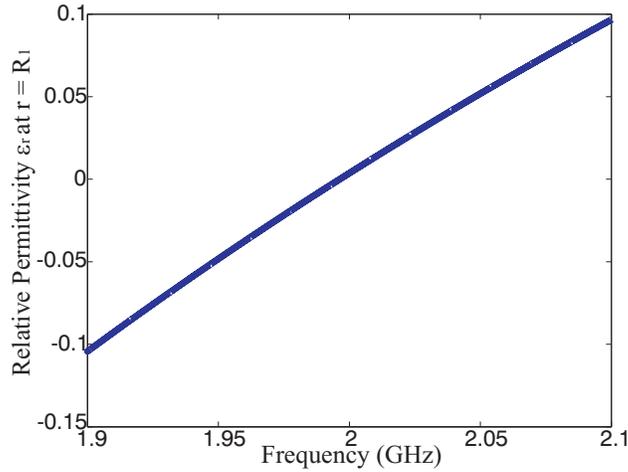}
\caption{Ideal cloaking material parameter $\varepsilon_{r}$ at the point $r=R_{1}$ of the cloak varying with frequency. Note that the values of $\varepsilon_{r}$ are always less than one and they can be negative.} \label{varyingepsr}
\end{figure}

Except from losses, which are directly affecting the cloak's performance, the cloaking material parameters are frequency-dispersive. For example, in Fig. \ref{varyingepsr} it can be seen how the value of $\varepsilon_{r}$
parameter at the inner radius of the cloaking device ($r=R_{1}$) is changing with a slight deviation from the central frequency of 2 GHz.
Hence, the cloak is functional only at a narrow frequency range. The FDTD method gives us the flexibility to easily perceive the bandwidth issues of the cloaking device because it is a time-domain numerical method. Again an FDTD modeling of the lossless cloaking device will be employed in order to investigate the bandwidth limitations of the cloak. The computation domain is the same as Fig. \ref{fdtddomainofcloaking} and the FDTD updating equations are the previously mentioned (\ref{discreteFaradaylaw}), (\ref{discreteAmperelaw}), (\ref{FDTDequationfinal}), (\ref{FDTDequationfinalEy}), (\ref{FDTDequationfinalHz}). The excitation for this case is a narrowband Gaussian pulse with a bandwidth between 1.65 GHz to 2.35 GHz and a central frequency of 2 GHz.

\begin{figure}[h]
\centering
\includegraphics[width=9cm]{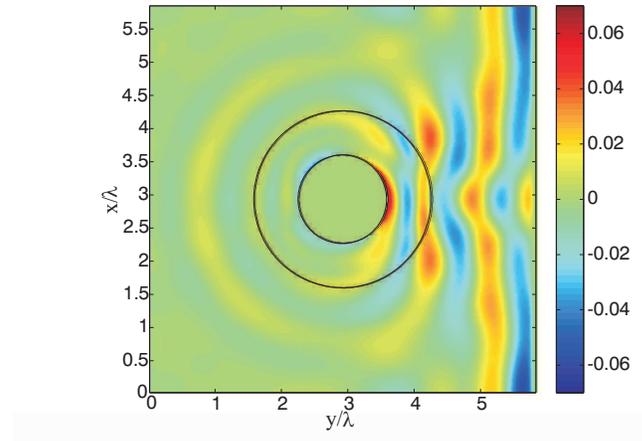}
\caption{A narrowband Gaussian pulse propagating from the left to the right side of the cloak. The pulse has a a bandwidth between 1.65GHz to 2.35GHz and a central frequency of 2GHz. The snapshot is taken when the pulse is recomposed at the right side of the cloak.} \label{gaussianpulseoutcloak}
\end{figure}

\begin{figure}[h]
\centering
\includegraphics[width=9cm]{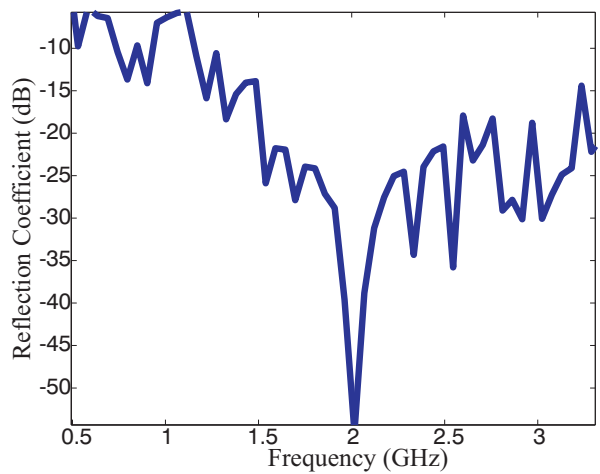}
\caption{Reflection coefficient of cloak in dB varying with frequency. The cloak is illuminated with a narrowband Gaussian pulse.} \label{reflectioncoefficient}
\end{figure}

\begin{figure}[h]
\centering
\includegraphics[width=9cm]{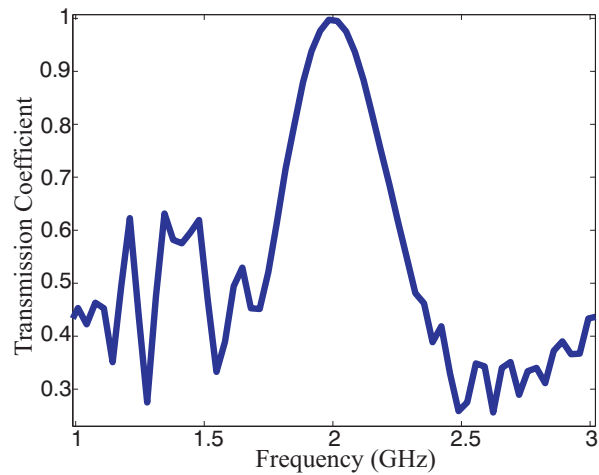}
\caption{Transmission coefficient of cloak varying with frequency. The cloak is illuminated with a narrowband Gaussian pulse.} \label{transmissioncoefficient}
\end{figure}

In Fig. \ref{gaussianpulseoutcloak} the Gaussian pulse can be observed after it has been propagated through the cloaking device.
It is obvious that there are reflections and the pulse trajectory is not recomposed correctly. However, the bending of the electromagnetic pulse inside the device is similar with the ideal cloak. The reflection coefficient of the cloaking device is calculated in order to measure the backscattering of the structure. The magnetic field values $H_{z}$ are averaged at a parallel to the x-axis line close to the plane wave source and the excitation pulse is isolated from the reflected signal. Furthermore, the transmission coefficient is measured with the same technique of averaging the field values at a line close to the right side PML. The computed reflection and transmission coefficients can be seen in Figs. \ref{reflectioncoefficient}, \ref{transmissioncoefficient} respectively.
It can be concluded that the cloak works properly (no reflections and totally transmission of the field) in a very narrow frequency range, ideally at a single frequency which is the central frequency of 2 GHz. However, it is interesting that the device can operate with a tolerable percentage of reflections and a half fraction of transmitted signal at a wider frequency range.

\section{Conclusions}
In this paper, a novel radial-dependent dispersive FDTD technique is proposed to model a lossy cloaking device. The cloaking material parameters are mapped to the dispersive Drude model and the constitutive equations are discretised in space and time. Both lossless and lossy electromagnetic cloaks have been investigated and the FDTD simulation results are in good agreement with those from the theoretical analysis and the frequency domain numerical modeling of the cloaking structure. From the FDTD numerical modeling, it can be concluded that the cloaking structure is sensitive to losses. Moreover, it can be perfectly ``invisible" only at a very narrow frequency range. However, it would be very interesting to study how lossy cloaks behave over a wide frequency range. Novel microwave absorbers and antenna structures may be developed based on the concept of coordinate transformation used in electromagnetic cloaking.

\balance

\end{document}